\documentclass[manuscript]{aastex}
\newcommand{\HIIR}{H{\sc ii} region}
\newcommand{\myemail}{tb@star.ucl.ac.uk}
\shorttitle{RDI and Triggered Star Formation}
\shortauthors{Bisbas et al.}
\usepackage{psfrag}
\begin{document}

\title{Radiation Driven Implosion and Triggered Star Formation}

\author{Thomas G. Bisbas \altaffilmark{1,2,3}, Richard W\"unsch\altaffilmark{2}, Anthony P. Whitworth\altaffilmark{3}, David A. Hubber\altaffilmark{4,5}, and Stefanie Walch\altaffilmark{3}}

\altaffiltext{1}{Department of Physics \& Astronomy, University College London, London WC1E 6BT, UK. (\myemail)}
\altaffiltext{2}{Astronomical Institute, Academy of Sciences of the Czech Republic, Bo\v{c}n\'{i} II 1401, 141 31 Prague, Czech Republic.}
\altaffiltext{3}{School of Physics \& Astronomy, Cardiff University, Cardiff CF24 3AA, UK.}
\altaffiltext{4}{Department of Physics \& Astronomy, University of Sheffield, Sheffield S3 7RH, UK.}
\altaffiltext{5}{School of Physics \& Astronomy, University of Leeds, Leeds LS2 9JT, UK.}

\begin{abstract}
We present simulations of initially stable isothermal clouds exposed to ionising radiation from a discrete external source, and identify the conditions that lead to radiatively driven implosion and star formation. We use the Smoothed Particle Hydrodynamics code SEREN and an HEALPix-based photo-ionisation algorithm to simulate the propagation of the ionising radiation and the resulting dynamical evolution of the cloud. We find that the incident ionising flux, $\Phi_{_{\rm LyC}}$, is the critical parameter determining the cloud evolution. At moderate fluxes, a large fraction of the cloud mass is converted into stars. As the flux is increased, the fraction of the cloud mass that is converted into stars and the mean masses of the individual stars both decrease. Very high fluxes simply disperse the cloud. Newly-formed stars tend to be concentrated along the central axis of the cloud (i.e. the axis pointing in the direction of the incident flux). For given cloud parameters, the time, $t_{_\star}$, at which star formation starts is proportional to $\Phi_{_{\rm LyC}}^{-1/3}$. The pattern of star formation found in the simulations is similar to that observed in bright-rimmed clouds.
\end{abstract}

\keywords{hydrodynamics --- methods: numerical --- stars: formation --- H II regions}

\section{Introduction}\label{SEC:INTRO}%

When an expanding {\HIIR} overruns a pre-existing cloud, it drives an ionisation front and a shock wave into the cloud \citep{San82, Ber89, Lef94}. As a consequence, the inner parts of the cloud are compressed, and may become gravitationally unstable, collapsing to form new stars. At the same time, the outer parts are ``boiled off'' by the ionisation front, and the rest of the neutral gas is stretched out into a cometary globule, with a dense head surrounding the newly-formed stars, a bright rim pointing towards the ionising star, and a cometary tail pointing away from the ionising star.

There are many observations suggesting that star formation has been triggered inside cometary globules and bright-rimmed clouds \citep[e.g.][]{Lef95,Lef97,Sug99,Sug00,Sug02,Wal02,Mor08,Smi10}. However, the extent to which star formation has actually been triggered by the shock front that preceeds a D-type ionisation front -- as opposed to simply being revealed by the dispersal of residual gas -- is not always clear \citep[e.g.][]{Dal07,Ind07,Gua10,Elm11}. There is some evidence that the young stellar objects deeply embedded inside a bright-rimmed cloud are the youngest ones, whilst those close to the rim of the cloud or distributed inside the expanding {\HIIR} are older \citep[e.g.][]{Ike08,Cha09}; this implies sequential triggering. In addition, \citet{Sug99,Sug00} report that the young stellar objects detected inside bright-rimmed clouds tend to lie close to the line joining the centre of mass of the cloud to the ionising star.

Simulations of the interaction of ionising radiation with self-gravitating clouds have been presented by various authors. \citet{Kes03} show that including self-gravity leads to greater compression, and may therefore trigger star formation. \citet{Gri09} show that stable Bonnor-Ebert spheres can be driven into gravitational collapse, and that the final mass and age of the collapsed core depend on the incident flux of ionising radiation. \citet{Mia09} find that the morphological evolution of a cloud is very sensitive to the relative importance of self-gravity. \citet{Gri10} and \citet{Mac10} have evaluated the roles played by turbulence and thermal instability in the formation of pillar-like structures. \citet{Hen09} and \citet{Art11} have explored the way in which magnetic fields influence the evolution of an {\HIIR}, and conclude that they may inhibit fragmentation of the neutral gas swept up by the expanding {\HIIR}; their simulations include both a proper treatment of the thermal microphysics and detailed predictions of diagnostic line emission, but not self-gravity, so they are focussing on different issues from us. As noted by \citet{Deh05} ``no model explains \emph{where} star formation takes place (in the core or at its periphery) or \emph{when} (during the maximum compression phase, or earlier)''.

In this paper we perform simulations of pre-existing clouds exposed to ionising radiation from a discrete source, and identify the circumstances that lead to triggered star formation. The paper is organized as follows. In Section \ref{SEC:ICS} we define the initial conditions. In Section \ref{SEC:NUM} we describe the numerical methods. In Section \ref{SEC:RES} we present and analyse the results. We summarize our main conclusions in Section \ref{SEC:CONC}.

\section{Initial conditions}\label{SEC:ICS}%

\begin{figure}[h]
\centering
\psfrag{q1}{$D_{_{\rm CLOUD}}$}\psfrag{q2}{$R_{_{\rm CLOUD}}$}\psfrag{q3}{Ionisation Front}
\includegraphics[width=0.24\textwidth]{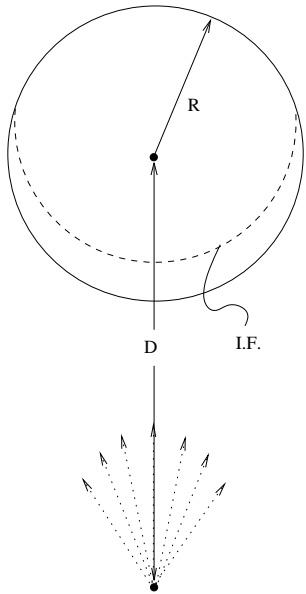}
\caption{Configuration of a cloud of radius $R_{_{\rm CLOUD}}$ that is ionised by an external star (solid black dot). The cloud is located at distance $D_{_{\rm CLOUD}}$ from the star. The dashed curve inside the cloud demarks the position of the ionisation front (I.F.) at the beginning of the simulation.}
\label{FIG:SKETCH}
\end{figure}

We define the initial conditions with reference to two coordinate systems, as illustrated schematically in Figure~\ref{FIG:SKETCH}.

The first co-ordinate system is centred on the ionising star. Distances from the ionising star are designated by the variable $D$. In particular, the position of the ionisation front in the direction of the unit vector $\hat{n}$ is given by $D_{_{\rm IF}}(\hat{\bf n})\,\hat{\bf n}$, and the centre of the cloud is initially at position $D_{_{\rm CLOUD}}\,\hat{\bf k}$, where $\hat{\bf k}$ is a unit vector along the $z$-axis (see Figure~\ref{FIG:SKETCH}). The ionising star is characterised by the rate, $\dot{\cal N}_{_{\rm LyC}}$, at which it emits hydrogen-ionising photons.

The second co-ordinate system is centred on the initial centre of mass of the cloud. Distances from this point are designated by the variable $R$. The fiducial cloud is modeled as a Bonnor-Ebert sphere, i.e. an equilibrium isothermal sphere \citep{Bon56, Ebe57}. It can therefore be characterised by its mass, $M_{_{\rm CLOUD}}=5\,{\rm M}_{_\odot}$, its isothermal sound speed, $a_{_{\rm O}}=0.2\,{\rm km}\,{\rm s}^{-1}$ (corresponding to molecular gas with fraction by mass of hydrogen $X=0.7$ and temperature $T=10\,{\rm K}$), and its Bonnor-Ebert parameter, $\xi_{_{\rm BE}}=4$ (the dimensionless radius). By choosing $\xi_{_{\rm BE}}<6.451$ we ensure that the cloud is initially stable, and therefore that its subsequent collapse is due to the incident ionising radiation. With this choice of parameters, the fiducial cloud initially has central density and radius given by
\begin{eqnarray}
\rho_{_{\rm CENTRE}}&=&\frac{a_{_{\rm O}}^6\,\mu^2\left(\xi_{_{\rm BE}}\right)}{4\,\pi\,G^3\,M_{_{\rm CLOUD}}^2}\;\,\simeq\;\, 10^{-20}\,{\rm g}\,{\rm cm}^{-3}\,,\\
R_{_{\rm CLOUD}}\,&=&\frac{G\,M_{_{\rm CLOUD}}\,\xi_{_{\rm BE}}}{a_{_{\rm O}}^2\,\mu\left(\xi_{_{\rm BE}}\right)}\;\;\;\simeq\;\,0.3\,{\rm pc}\,,
\end{eqnarray}
where $\mu\left(\xi_{_{\rm BE}}\right)$ is the dimensionless mass obtained by solving the Isothermal Equation \citep{Cha39, Cha49}.

The ionisation routine is designed to treat the $D$-type expansion of an {\HIIR}. Therefore each simulation starts when the ionisation front switches from $R$-type to $D$-type \citep{Kah54}. We assume that at this juncture the matter has not had time to move. On Figure~\ref{FIG:SKETCH} the ionisation front is denoted by a dashed line.

We have performed three suites of simulations. The first suite explores the consequences of changing $\dot{\cal N}_{_{\rm LyC}}$, and hence the flux of ionising photons, $\Phi_{_{\rm LyC}}=\dot{\cal N}_{_{\rm LyC}}/4\pi D_{_{\rm CLOUD}}^2$, incident on the fiducial cloud. In the second suite, we explore the consequences of changing $M_{_{\rm CLOUD}}$, keeping $a_{_{\rm O}}$, $\xi_{_{\rm BE}}$ and $R_{_{\rm CLOUD}}/D_{_{\rm CLOUD}}$ fixed, with a view to determining the maximum ionising flux that triggers star formation. In the third suite we explore the effect of numerical noise, by invoking different random  -- but smooth -- initial distributions of SPH particles. In all simulations, the cloud is placed at distance $D_{_{\rm CLOUD}}=10 R_{_{\rm CLOUD}}$ from the ionising star, so that the divergence of the ionising radiation impinging on the cloud, measured by $R_{_{\rm CLOUD}}/D_{_{\rm CLOUD}}=0.1$, is invariant.

\section{Numerical methods}\label{SEC:NUM}%

\subsection{SPH Code}%

We use the Smoothed Particle Hydrodynamics (SPH) code SEREN, which has been designed to investigate star and planet formation problems, and extensively tested on a wide range of standard problems, as described in \citet{Hub11} and on the SEREN webpage\footnote{http://www.astro.group.shef.ac.uk/seren}. In the present work we use the following SEREN options: standard SPH \citep{Mon92} with a fixed (and exact) number of neighbours ${\cal N}_{_{\rm NEIB}}=50$; standard artificial viscosity with $\alpha_{_{\rm AV}}=1$, $\beta_{_{\rm AV}}=2$ and no switches; an algorithm based on HEALPix\footnote{http://healpix.jpl.nasa.gov/} for following ionising radiation \citep[see Section \ref{SEC:IF} below; and][]{Bis09}; and a barotropic equation of state for the neutral gas (see Section \ref{SEC:THOD} below). The SPH equations of motion are solved with a second-order Leapfrog integrator, in conjunction with an hierarchical block time-stepping scheme. Gravitational forces are calculated using an octal-spatial tree \citep{Bar86}, with monopole and quadrupole terms and a standard geometric opening-angle criterion. The tree is also used to generate neighbour lists.

\subsection{Initial equilibrium cloud}%

The initial cloud configuration is created by first settling a cube of SPH particles of size $2\times 2\times 2$, using hydrostatic forces and periodic boundary conditions, but no self-gravity, to obtain a uniform-density glass-like distribution of particles. Next a unit-radius sphere is cut from this cube, the number of particles in the sphere, ${\cal N}_{_{\rm SPHERE}}$, is counted, and the particles are given equal masses $m_{_{\rm SPH}}=M_{_{\rm CLOUD}}/{\cal N}_{_{\rm SPHERE}}$. Finally the particle positions are stretched radially to reproduce a Bonnor-Ebert density profile; for a particle initially at radius $r$, we solve the equation
\begin{eqnarray}
\mu(\xi)&=&r^3\,\mu\left(\xi_{_{\rm BE}}\right)
\end{eqnarray}
for $\xi$, and then displace the particle radially according to
\begin{eqnarray}
r&\longrightarrow&\frac{R_{_{\rm CLOUD}}\,\xi}{\xi_{_{\rm BE}}}\,.
\end{eqnarray}

In all simulations we arrange that the mass of an SPH particle is $m_{_{\rm SPH}}\simeq 5\times 10^{-5}\,{\rm M}_{_\odot}$. Given the barotropic equation of state that we invoke (Eqn. \ref{EQN:BAROTROPIC}), the minimum Jeans mass is $M_{_{\rm MIN}}\sim 0.005\,{\rm M}_{_\odot}$, and is therefore just resolved with $\,\sim\! 2{\cal N}_{_{\rm NEIB}}\!\sim\!100\,$ SPH particles \citep{Bat97,Whi98,Hub06}.

\subsection{Propagation of the ionising radiation}\label{SEC:IF}%

The ionisation routine uses the HEALPix algorithm \citep{Gor05} to create an hierarchy of rays emanating radially from the ionising star \citep[cf.][]{Abe02}. In the immediate vicinity of the star there are only twelve rays, each at the centre of an approximately square element of solid angle $\Omega_{_0}=4\pi/12\simeq 1\,{\rm steradian}$. Each ray is then split adaptively into 4 child-rays, and the splitting is repeated recursively so that the separation between neighbouring rays is always less than one half of the local smoothing length (i.e. always less than one eighth of the diameter of the local SPH particles). Thus, after $\ell$ splittings each ray is at the centre of an approximately square element of solid angle $\Omega_{_\ell}=4^{-\ell}\,{\rm steradian}$, and is separated from its neighbours by an angle $\theta_{_\ell}\simeq 2^{-\ell}\,{\rm radian}$. In the present simulations there are up to eleven levels of splitting, so the rays on the finest level are $\sim 1.5\,{\rm arcmin}$ apart. At each step we randomise the orientation of the twelve rays on level 0 (and hence also the orientation of rays on higher levels), in order to avoid numerical artefacts \citep[cf.][]{Kru07}.

Along a ray with direction given by the unit vector $\hat{\bf n}$, the ionisation front is located at position $D_{_{\rm IF}}\!\left(\hat{\bf n}\right)\,\hat{\bf n}$, where
\begin{eqnarray}\label{EQN:IONBAL}
\int\limits_{D=0}^{D=D_{_{\rm IF}}}\,\rho^2(D\hat{\bf n})\;D^2\,dD&=&\frac{\dot{\cal N}_{_{\rm LyC}}\,m^2}{4\,\pi\,\alpha_{_{\rm B}}}\equiv I_{_{\rm MAX}}\,.
\end{eqnarray}
Here $\rho(D\hat{\bf n})$ is the mass-density at position $D\hat{\bf n}$, $m=m_{\rm p}/X$ is the mean mass associated with each hydrogen nucleus, $m_{\rm p}$ is the proton mass, $X=0.7$ is the fraction by mass of hydrogen, and $\alpha_{_{\rm B}}$ is the recombination coefficient into excited stages only \citep[i.e. we invoke the On-The-Spot Approximation;][]{Ost74}. The integral in Eqn. (\ref{EQN:IONBAL}) is evaluated using a second-order integration scheme, with an integration step that is one quarter of the local smoothing length (i.e. one sixteenth of the diameter of the local SPH particles).

Further details of the method are given in \citet{Bis09}.

\begin{figure}[h]
\centering
\includegraphics[width=0.8\textwidth,angle=0]{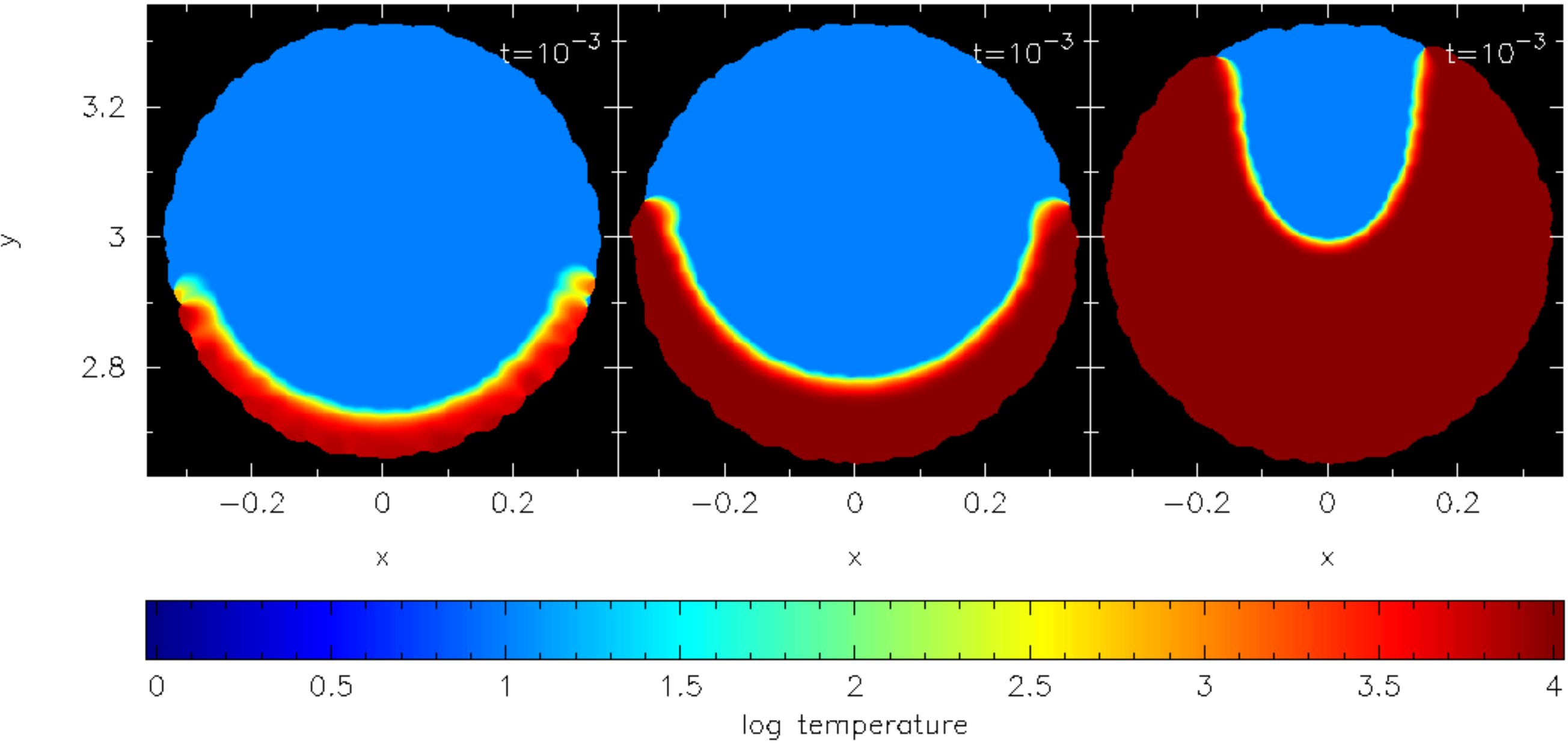}
\caption{False-colour maps of the initial gas temperature on slices through the middle of the fiducial cloud, for three different ionising fluxes. From left to right, $\Phi_{_{\rm LyC}}=10^9\,{\rm cm}^{-2}\,{\rm s}^{-1}$ (Simulation 1), $\Phi_{_{\rm LyC}}=3\times 10^{10}\,{\rm cm}^{-2}\,{\rm s}^{-1}$ (Simulation 4), and $\Phi_{_{\rm LyC}}=10^{12}\,{\rm cm}^{-2}\,{\rm s}^{-1}$ (Simulation 7). The colour bar gives $\log_{_{10}}(T/{\rm K})$, and the axes are in pc. The snapshots are taken after the first iteration of the code; the time on each map denotes to the moment when the ionisation front switches from $R$ to $D$ type.}
\label{FIG:TEMPMAP}
\end{figure}

\subsection{Thermodynamics}\label{SEC:THOD}%

\noindent{\sc Neutral gas.} The temperature, $T_{_{\rm N}}$, of the neutral gas is calculated using a barotropic equation of state,
\begin{equation}
\label{EQN:BAROTROPIC}
T_{_{\rm N}}(\rho)=T_{_{\rm O}}\left\{1+\left(\frac{\rho}{\rho_{_{\rm CRIT}}}\right)^{\gamma -1}\right\}\,,
\end{equation}
where $T_{_{\rm O}}=10\,{\rm K}$, $\rho_{_{\rm CRIT}}=10^{-13}\,{\rm g}\,{\rm cm}^{-3}$ and $\gamma =5/3$ is the ratio of specific heats. This equation of state is presumed to mimic the gross behaviour of protostellar gas: at low densities, $\rho\ll\rho_{_{\rm CRIT}}$, the gas tends to be optically thin to its own cooling radiation and therefore approximately isothermal at $T\sim 10\,{\rm K}$; at high densities, $\rho\gg\rho_{_{\rm CRIT}}$, the gas tends to be optically thick to its own cooling radiation and therefore heats up adiabatically. Since the gas dynamics is only followed to densities $\sim 10^{-11}\,{\rm g}\,{\rm cm}^{-3}$ (thereafter sink particles are introduced; see below), the temperature is always $T\la 200\,{\rm K}$. Therefore we can assume that $\gamma\sim 5/3$, even for H$_2$, because the rotational degrees of freedom of H$_2$ are only weakly excited.

\vspace{0.1cm}
\noindent{\sc Ionised gas.} The temperature, $T_{_{\rm I}}$, of the ionised gas is assumed to be $T_{_{\rm I}}=10^4\,{\rm K}$, except in the immediate vicinity of the ionisation front, where the temperature changes smoothly from $T_{_{\rm I}}$ to $T_{_{\rm N}}$ over a region with a width of two smoothing lengths \citep[i.e. half the diameter of an SPH particle; see][]{Bis09}. Figure~\ref{FIG:TEMPMAP} shows initial temperature cross-sections through the fiducial cloud for three representative ionising fluxes, at the time when the ionisation front switches from $R$ to $D$ type.

\subsection{Sink particles}%

We follow \citet{Bat95} in replacing the densest regions with sink particles. Specifically, if the density of SPH particle $i$ exceeds $\rho_{_{\rm SINK}}=10^{-11}\,{\rm g}\,{\rm cm}^{-3}$, then SPH particle $i$ and its neighbours are excised from the simulation and replaced by a sink particle having the same mass, centre of mass and momentum as the sum of the SPH particles it replaces. Any SPH particle which subsequently passes within $R_{_{\rm SINK}}\simeq 4\,{\rm AU}$ of a sink, and is gravitationally bound to it, is accreted by the sink, i.e. it is excised from the simulation and its mass and momentum are assimilated by the sink. The presumption is that the material in a sink will inevitably collapse to form a single star or a close multiple system. We therefore hereafter refer to sinks as stars. However, we do not take account of radiative feedback from these newly-formed stars.

The value of $\rho_{_{\rm SINK}}$ is chosen so that any protostellar condensation is already well into its Kelvin-Helmholtz contraction phase (and therefore well relaxed), but not yet into the second collapse phase, when it is replaced with a sink. With $\rho_{_{\rm SINK}}=10^{-11}\,{\rm g}\,{\rm cm}^{-3}$, the temperature inside a protostellar condensation has increased from $\sim 10\,{\rm K}$ to $\sim 200\,{\rm K}$, by the time it becomes a sink. The sink radius is not specified, but is calculated on-the-fly to ensure that a protostellar condensation that becomes a sink is resolved; this gives $R_{_{\rm SINK}}\simeq(3M_{_{\rm MIN}}/4\pi\rho_{_{\rm SINK}})^{1/3}\simeq 4\,{\rm AU}$.

\subsection{Magnetic fields}%

The SEREN code is not able to handle magnetic fields, and so our results must be interpreted as exploring what can happen when the magnetic field is dynamically unimportant. It will be important to explore this aspect of the problem, but even if a uniform magnetic field were assumed, it would necessarily introduce two further free parameters (field strength and direction). Previous studies of irradiated clouds embedded in a uniform magnetic field suggest that a sufficiently strong magnetic field may seriously affect the cloud evolution \citep[][]{Wil07,Hen09,Mac10}. Depending on the initial field orientation with respect to the direction of ionising photons, the cloud is either flattened (perpendicular orientation) or the radiative implosion may be prevented (parallel orientation). However, these findings were not confirmed by \citet{Art11} who studied formation and evolution of globules at the border of the {\HIIR} expanding into the turbulent, magnetised molecular cloud, and found that the magnetic effects are weak. Further studies are needed to resolve this discrepancy. A turbulent field and/or non-ideal effects would complicate matters still further.

\begin{table}
\begin{center}
\caption{The ID, output of ionising photons from the central star, $\dot{\cal N}_{_{\rm Lyc}}$, and ionising flux incident on the cloud, $\Phi_{_{\rm LyC}}$, for the simulations involving the fiducial cloud.}
\label{TAB:SUITE1}\centering\begin{tabular}{ccccccccc}
\tableline\tableline
Simulation ID & 1 & 2 & 3 & 4 & 5 & 6 & 7 & 8 \\\tableline
$\dot{\cal N}_{_{\rm Lyc}}/{\rm s}^{-1}$ & $\stackrel{}{10^{48}}$ & $3\times10^{48}$ & $10^{49}$ & $3\times10^{49}$ & $10^{50}$ & $3\times10^{50}$ & $10^{51}$ & $3\times10^{51}$ \\
$\Phi_{_{\rm Lyc}}/{\rm cm}^{-2}{\rm s}^{-1}$ & $10^{9}$ & $3\times10^{9}$ & $10^{10}$ & $3\times10^{10}$ & $10^{11}$ & $3\times10^{11}$ & $10^{12}$ & $3\times10^{12}$ \\
\tableline
\end{tabular}
\end{center}
\end{table}

\section{Results}\label{SEC:RES}%

\subsection{The fiducial cloud exposed to different ionising fluxes}\label{SEC:SUITE1}%

In the first suite of simulations, we treat the fiducial cloud (with mass $M_{_{\rm CLOUD}}=5\,{\rm M}_{_\odot}$, isothermal sound speed $a_{_{\rm O}}=0.2\,{\rm km}\,{\rm s}^{-1}$ and Bonnor-Ebert parameter $\xi_{_{\rm BE}}=4$, hence radius $R_{_{\rm CLOUD}}=0.3\,{\rm pc}$; we place it at distance $D_{_{\rm CLOUD}}=10R_{_{\rm CLOUD}}=3\,{\rm pc}$ from the ionising star; and we simply vary the ionising output of the star,  $\dot{\cal N}_{_{\rm LyC}}$, and hence the flux of ionising photons incident on the cloud, $\Phi_{_{\rm LyC}}$. Table \ref{TAB:SUITE1} gives the ID, $\dot{\cal N}_{_{\rm LyC}}$ and $\Phi_{_{\rm LyC}}$ for each simulation in this suite.

\subsubsection{Overview}%

In all simulations, the ionising radiation erodes the cloud from one side, and the ionised gas flows away into the surrounding space. There is a range of ionising fluxes, that trigger star formation in the cloud; for the fiducial cloud this range is $10^9\,{\rm cm}^{-2}\,{\rm s}^{-1}\la\Phi_{_{\rm LyC}}\la 3\times 10^{11}\,{\rm cm}^{-2}\,{\rm s}^{-1}$. If the ionising flux is very low, the ionisation front is R-type, and the cloud is just slowly eroded. If the ionising flux is $\ga 10^9\,{\rm cm}^{-2}\,{\rm s}^{-1}$, the ionisation front becomes D-type, and is preceded by a shock front, which propagates into the cloud compressing it. There is then a competition between erosion of the cloud by the ionisation front and collapse of the cloud triggered by the shock front. Provided the ionising flux is not too large, $\Phi_{_{\rm LyC}}\la 3\times 10^{11}\,{\rm cm}^{-2}\,{\rm s}^{-1}$, at least some of the cloud ends up in newly-formed stars. However, if the ionising flux is very large, the cloud is rapidly dispersed, with no star formation. All simulations are followed until all the cloud gas is either in newly-formed stars, or has been ionised and dispersed. The time at which the first star forms is denoted by $t_{_\star}$. We limit further discussion to the cases in which star formation occurs, namely those with IDs 1 to 6 in Table \ref{TAB:SUITE1}.

\subsubsection{Phenomenology of star formation for low ionising fluxes}%

\begin{figure}[h]
\centering
\includegraphics[width=0.8\textwidth,angle=-90]{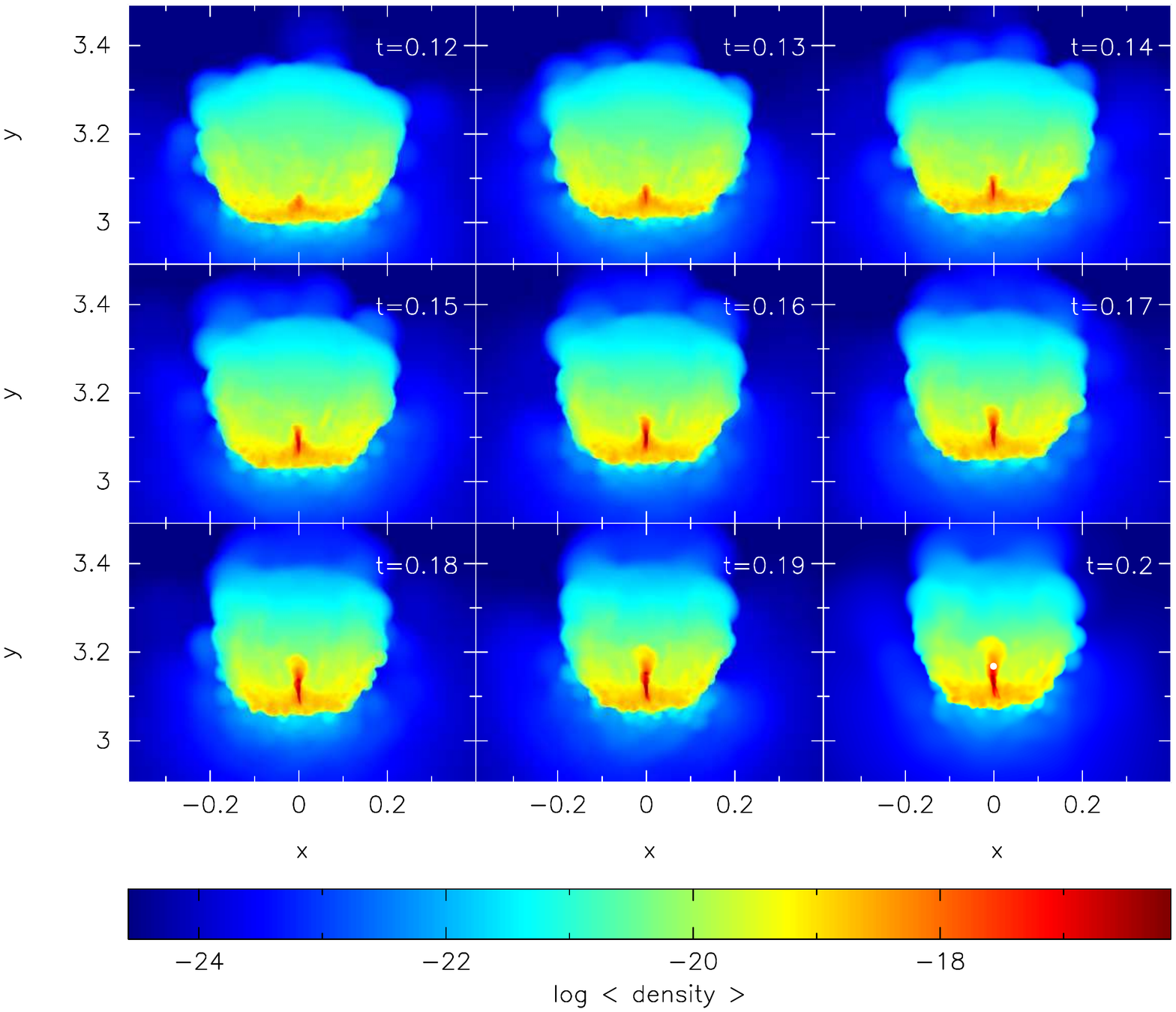}
\caption{Time sequence of surface-density images showing the evolution of the fiducial cloud when it is exposed to a relatively low ionising flux (Simulation 1, $\Phi_{_{\rm LyC}}=10^9\,{\rm cm}^{-2}\,{\rm s}^{-1}$). Note the overall broad \textbf{U}-shape of the ionisation front bounding the exposed side of the cloud, and the dense filament forming down the centre of the cloud. $t/{\rm Myr}$ is given in the top righthand corner of each frame. The colour bar gives $\log_{_{10}}(\Sigma/{\rm g}\,{\rm cm}^{-2})$, where $\Sigma$ is the surface-density, and the axes are in pc.}
\label{FIG:LOWSEQ}
\end{figure}

If the incident ionising flux is at the low end of the range for triggered star formation, the shock front driven into the exposed side of the cloud is weak, and it sweeps up matter rather slowly. Consequently, long before the shock front reaches the centre of the cloud, the pancake-shaped layer of swept-up gas becomes sufficiently massive to contract laterally due to its own self-gravity. As the material in the layer converges on the axis of symmetry, the increase in pressure deflects the inward flow towards the centre of the cloud, creating a dense filament down the axis of symmetry. It is in this filament that star formation first occurs, usually at a point well behind the ionisation front and towards the centre of the cloud. For these relatively low ionising fluxes, the ionisation front on the exposed side of the cloud develops a \textbf{U}-shape structure. Figure~\ref{FIG:LOWSEQ} is a time sequence of column-density images for Simulation 1, illustrating these morphological features, and terminating at the moment the first star forms.

\subsubsection{Phenomenology of star formation for high ionising fluxes}%

\begin{figure}[h]
\centering
\includegraphics[width=0.8\textwidth,angle=-90]{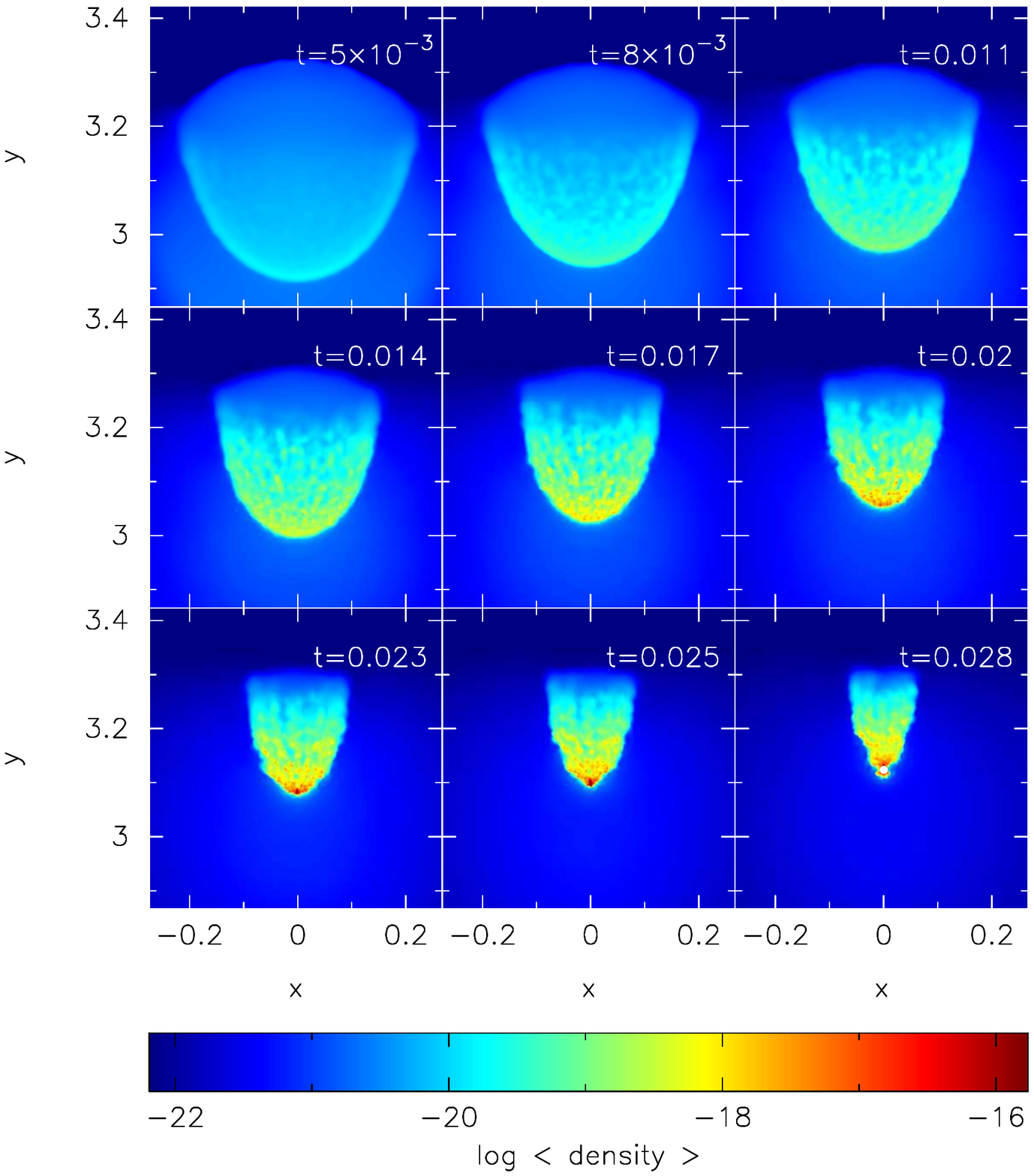}
\caption{Time sequence of surface-density images showing the evolution of the fiducial cloud when it is exposed to a relatively high ionising flux (Simulation 6, $\Phi_{_{\rm LyC}}=3\times 10^{11}\,{\rm cm}^{-2}\,{\rm s}^{-1}$). Note the overall narrow \textbf{V}-shape of the ionisation front bounding the exposed side of the cloud, and the sharp density peak at the apex of the cloud pointing towards the ionising star. $t/{\rm Myr}$ is given in the top righthand corner of each frame. The colour bar gives $\log_{_{10}}(\Sigma/{\rm g}\,{\rm cm}^{-2})$, where $\Sigma$ is the surface-density, and the axes are in pc.}
\label{FIG:HIGHSEQ}
\end{figure}

If the incident ionising flux is at the high end of the range for triggered star formation, the shock front driven into the exposed side of the cloud is strong, and it sweeps up matter rather rapidly. As a result the swept-up layer becomes sufficiently massive to fragment, before it has had time to undergo much lateral contraction under its own self-gravity, and therefore before there has been time to form an axial filament. Consequently, star formation first occurs near the centre of the swept-up layer, and close to the ionisation front. For these relatively high ionising fluxes, the ionisation front on the exposed side of the cloud develops a \textbf{V}-shape structure. Figure~\ref{FIG:HIGHSEQ} is a time sequence of column-density images for Simulation 6, illustrating these morphological features, and terminating at the moment the first star forms.

\subsubsection{Formation time of first star}%

\begin{figure}[h]
\centering
\includegraphics[width=0.5\textwidth,angle=-90]{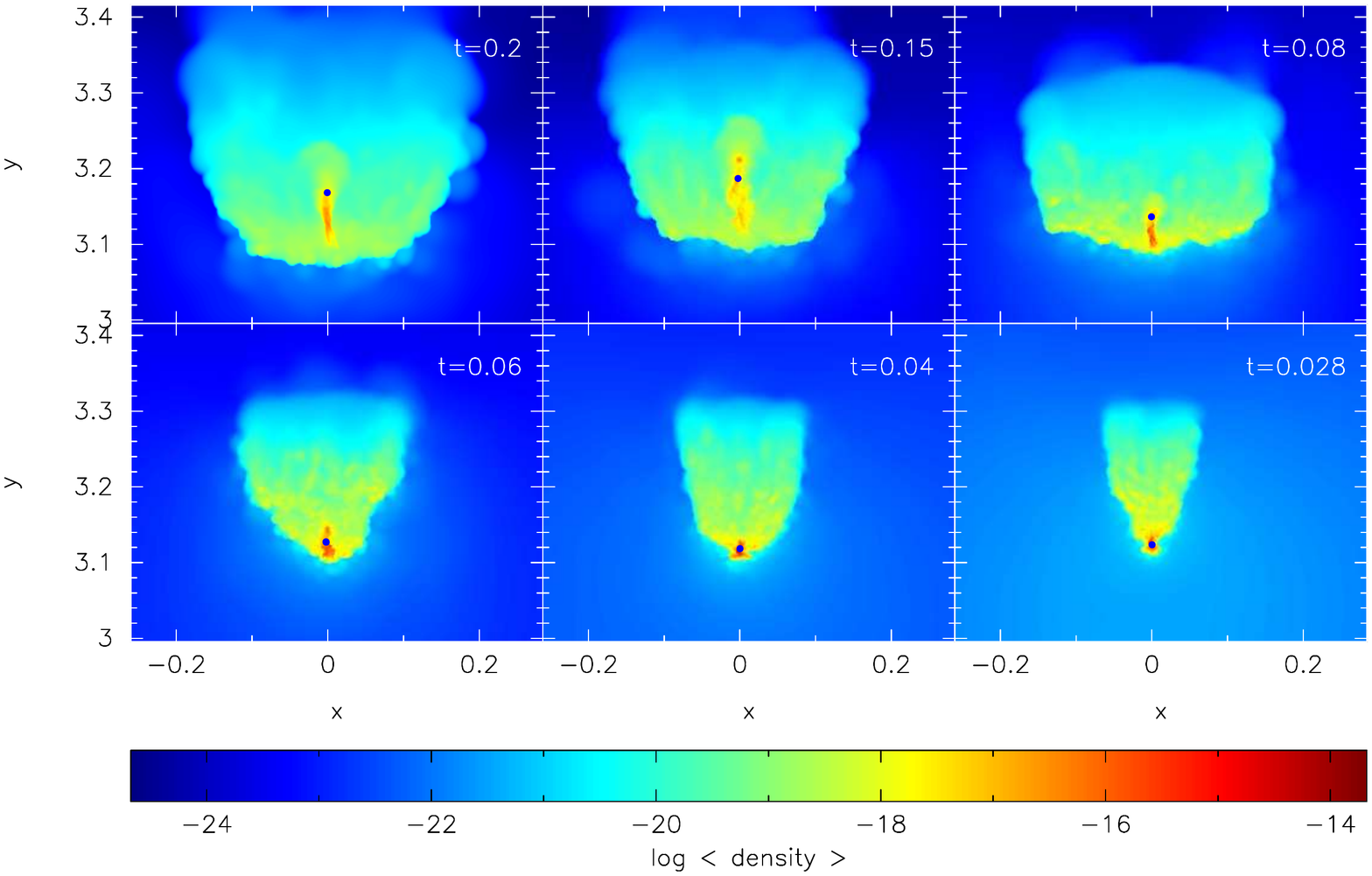}
\caption{Surface-density images of the fiducial cloud at the moment the first star forms, $t_{_\star}$, for different ionising fluxes, as given in Table \ref{TAB:SUITE1}: top row, left to right, Simulations 1, 2 and 3; bottom row, left to right, Simulations 4, 5 and 6. $t_{_\star}/{\rm Myr}$ is given in the top right corner of each frame. The location of the first star is marked with a dark blue dot. The colour bar gives $\log_{_{10}}(\Sigma/{\rm g}\,{\rm cm}^{-2})$, where $\Sigma$ is the surface-density, and the axes are in pc.}
\label{FIG:SUITE1}
\end{figure}

\begin{figure}[h]
\centering
\includegraphics[width=0.8\textwidth]{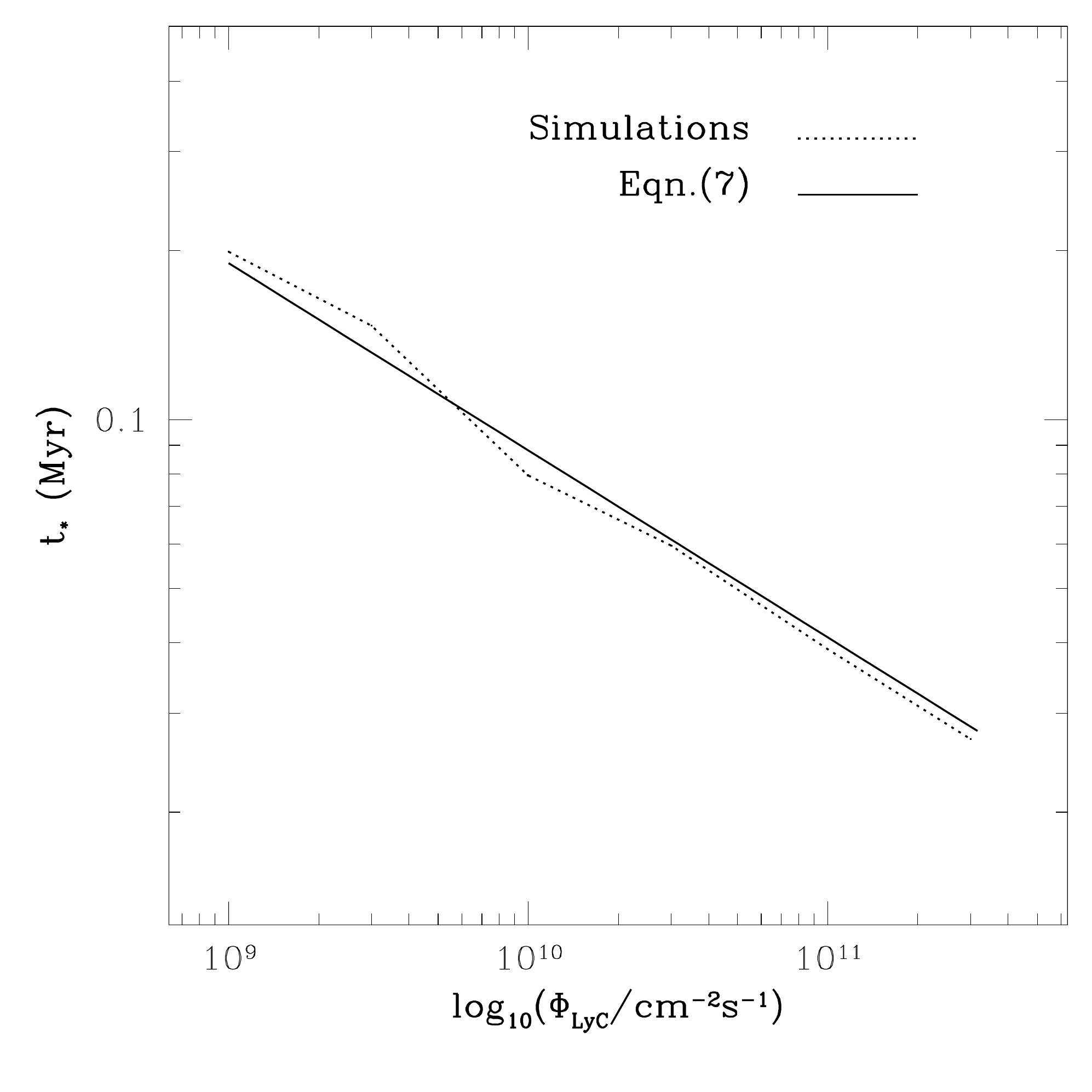}
\caption{Logarithmic plot of the time at which the first star forms, $t_{_\star}$ against the ionising flux, $\Phi_{_{\rm LyC}}$. The solid line is the best fit power-law given by Eqn. (\ref{EQN:TSTAR}).}
\label{FIG:TSTAR}
\end{figure}

Figure~\ref{FIG:SUITE1} shows column-density images of all the simulations involving the fiducial cloud and resulting in star formation, at the moment star formation occurs, $t_{_\star}$. This sequence demonstrates how, as the ionising flux is increased, the length of the central filament shortens, the location of star formation moves closer to the ionisation front, and the shape of the ionisation front morphs from a broad {\bf U}-shape to a narrower {\bf V}-shape. As noted by \citet{Gri09}, star formation occurs earlier for higher ionising fluxes. Figure~\ref{FIG:TSTAR} is a plot of $t_{_\star}$ against $\Phi_{_{\rm LyC}}$, for the fiducial cloud. The dashed line shows the numerical results, and the solid line shows the best power-law fit to these results,
\begin{eqnarray}\label{EQN:TSTAR}
t_{\star}&\simeq&0.19\,{\rm Myr}\,\left(\frac{\Phi_{_{\rm LyC}}}{10^9\,{\rm cm}^{-2}\,{\rm s}^{-1}}\right)^{-1/3}\,.
\end{eqnarray}

\subsubsection{Cloud morphology at $t_{_\star}$}%

\begin{figure}
\centering
\includegraphics[width=0.37\textwidth]{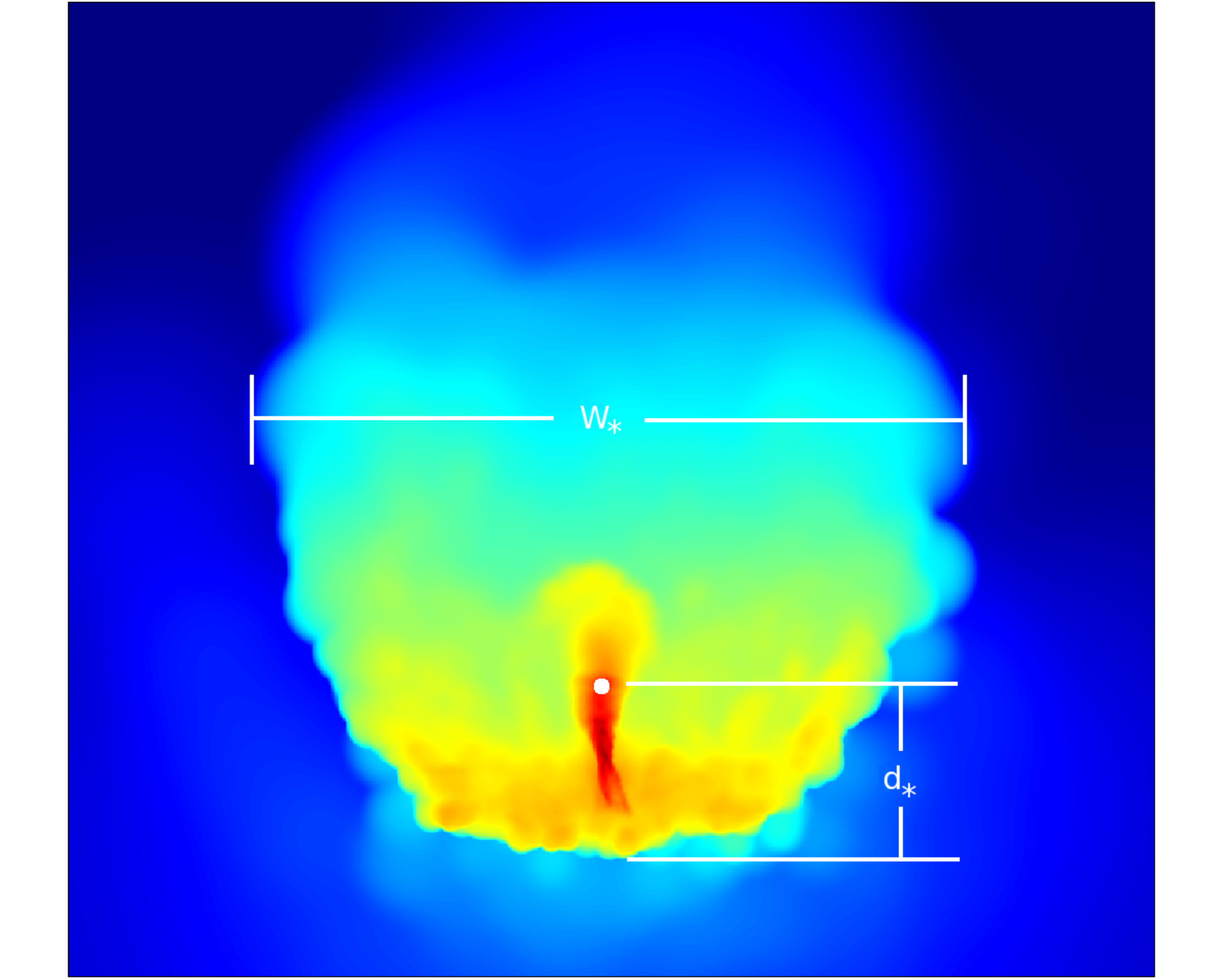}
\includegraphics[width=0.3\textwidth]{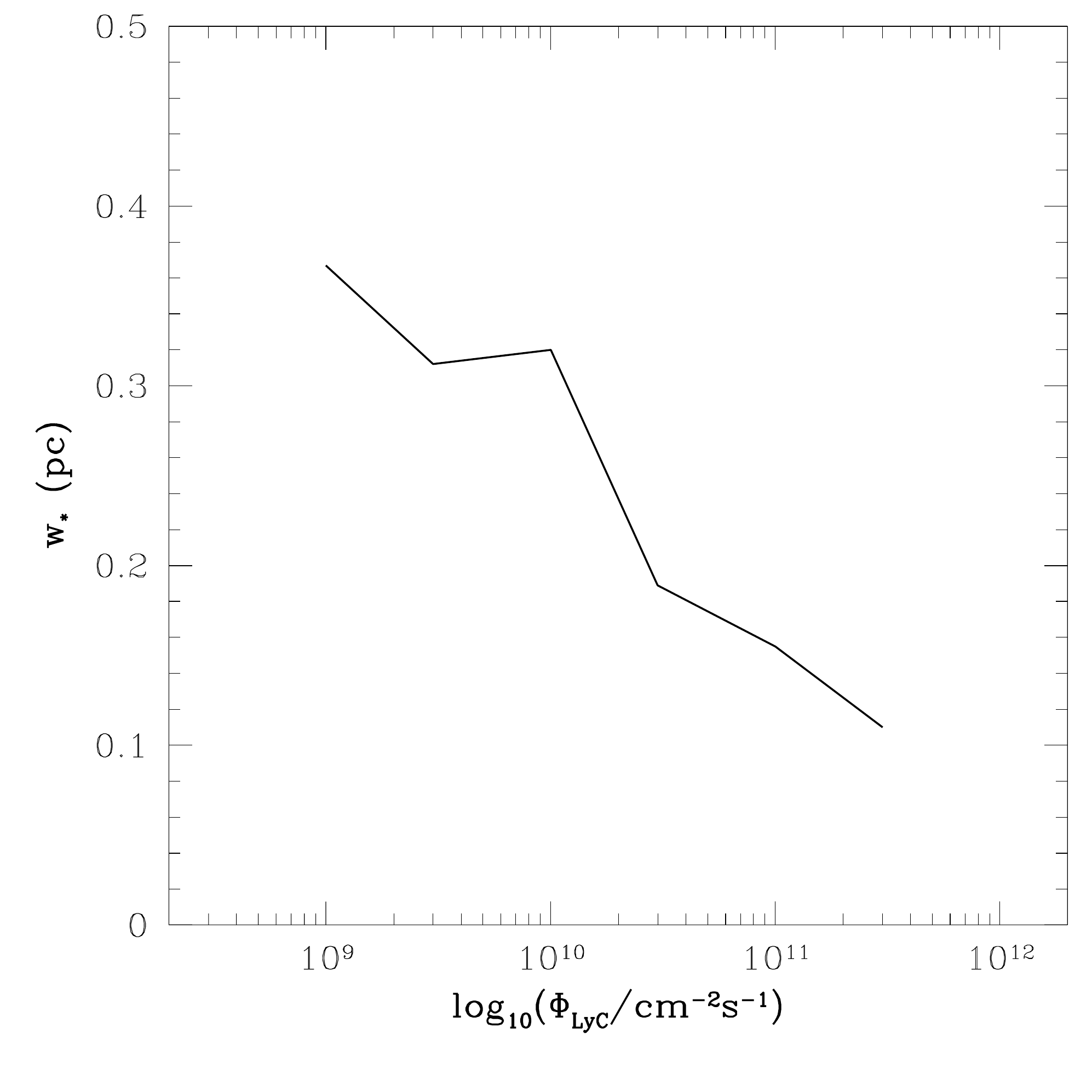}
\includegraphics[width=0.3\textwidth]{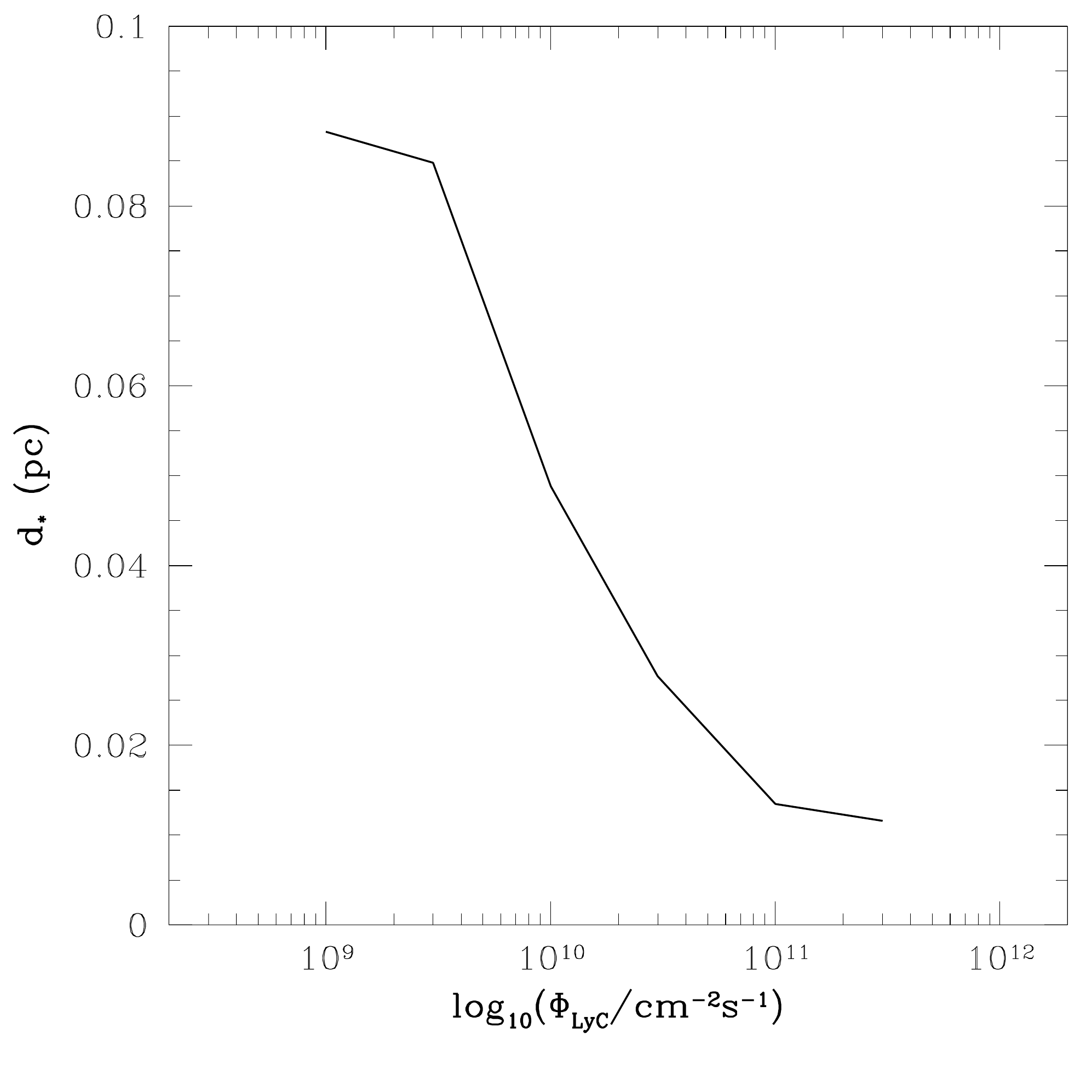}
\caption{\emph{Left panel}: Snapshot of Simulation 1 at $t_{_\star}$, illustrating schematically the definitions of $w_{_\star}$ and $d_{_\star}$. \emph{Middle panel}: $w_{_\star}$ as a function of $\log_{_{10}}(\Phi_{_{\rm LyC}})$ for Simulations 1 to 6. \emph{Right panel}: $d_{_\star}$ as a function of $\log_{_{10}}(\Phi_{_{\rm LyC}})$ for Simulations 1 to 6.}
\label{FIG:MORPH}
\end{figure}

The morphological evolution of the cloud and the location of star formation can be quantified coarsely by defining two metrics at $t_{_\star}$ (see lefthand panel of Figure~\ref{FIG:MORPH}).

The first metric is the width, $w_{_\star}$ of the cloud at $t_{_\star}$, i.e. the maximum extent of the neutral gas perpendicular to the direction of the ionising star. The variation of $w_{_\star}$ with $\log_{_{10}}(\Phi_{_{\rm LyC}})$ is shown in the middle panel of Figure~\ref{FIG:MORPH}. On average, $w_{_\star}$ decreases with increasing $\Phi_{_{\rm LyC}}$, from $w_{_\star}>0.3\,{\rm pc}$ at the low fluxes corresponding to broad {\bf U}-shaped ionisation fronts, to $w_{_\star}<0.2\,{\rm pc}$ at the high fluxes corresponding to narrow {\bf V}-shaped ionisation fronts.

The second metric is the distance, $d_{_\star}$ from the ionisation front to the first star, at $t_{_\star}$. The variation of $d_{_\star}$ with $\log_{_{10}}(\Phi_{_{\rm LyC}})$ is shown in the righthand panel of Figure~\ref{FIG:MORPH}. $d_{_\star}$ decreases monotonically with increasing $\Phi_{_{\rm LyC}}$, from $d_{_\star}>0.08\,{\rm pc}$ at low $\Phi_{_{\rm LyC}}$, to $d_{_\star}<0.02\,{\rm pc}$ at high $\Phi_{_{\rm LyC}}$.

\subsubsection{Star formation after $t_{_\star}$}\label{SEC:EGG}%

\begin{figure}[h]
\centering
\includegraphics[width=0.8\textwidth]{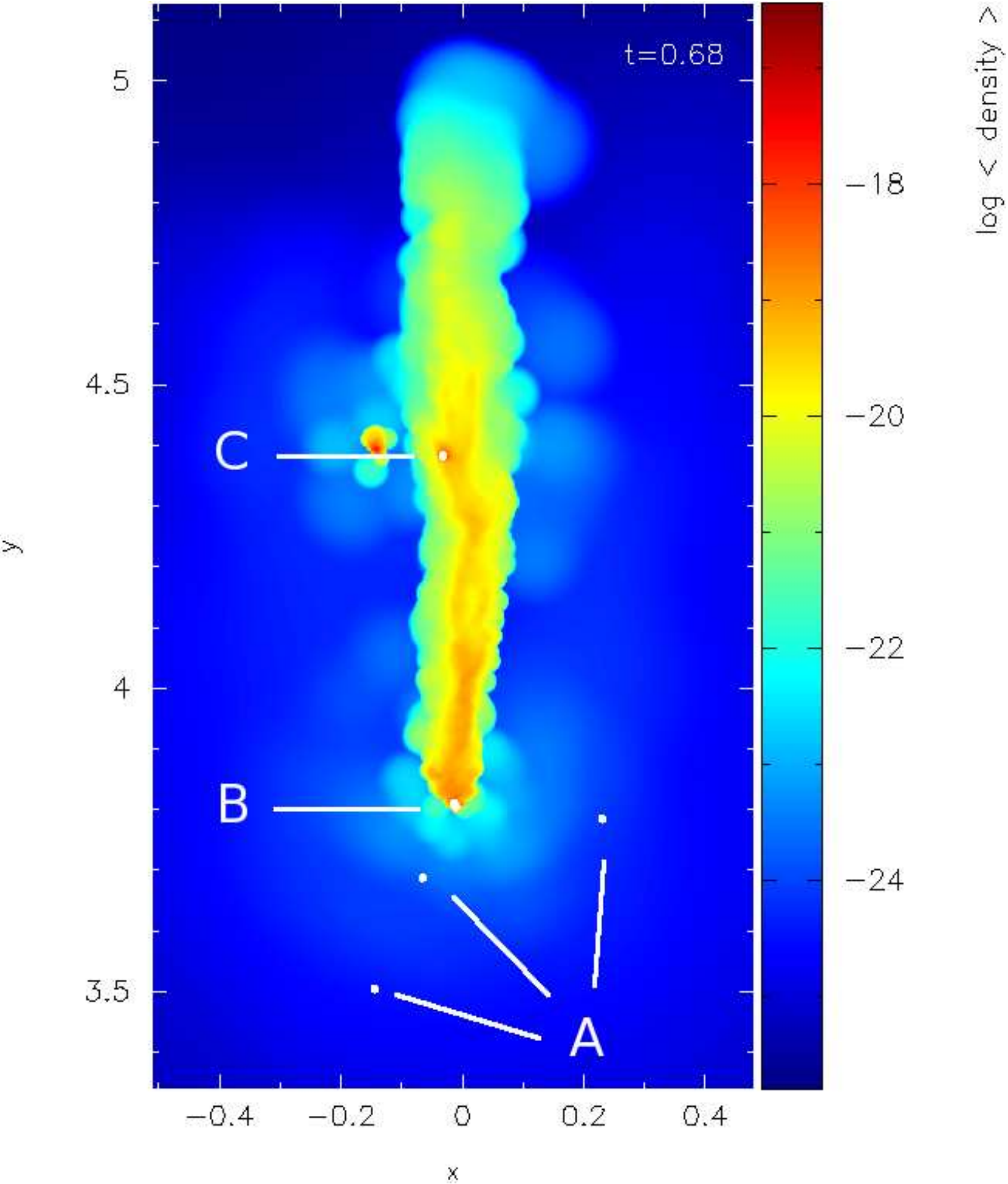}
\caption{Surface-density image of Simulation 2 at $t=0.68\,{\rm Myr}$. The colour bar gives $\log_{_{10}}(\Sigma/{\rm g}\,{\rm cm}^{-2})$, where $\Sigma$ is the surface-density, and the axes are in pc. See Section \ref{SEC:EGG} for discussion.}
\label{FIG:EGG}
\end{figure}

After $t_{_\star}$, further star formation is usually confined to the region between the first star and the ionisation front. Once the shock front has passed, any neutral gas that has not yet been incorporated into stars is overpressured, and tends to re-expand, as noted previously by \citet{San82, Ber89, Lef94, Gri09, Mia09}.

For low ionising fluxes, the expanding neutral material forms a single massive pillar protruding into the \HIIR, with a dense head pointing towards the ionising star; this is where most of the additional stars are formed. Because of their location, they are quickly overrun by the ionisation front. This terminates accretion, and therefore they tend to have low masses. The external appearance of these stars as they emerge into the {\HIIR} is very reminiscent of the Evaporating Gaseous Globules (EGGs) seen in M16 \citep{Hes96} and elsewhere.

Figure~\ref{FIG:EGG} illustrates the remnants of the single pillar formed in Simulation 2, at $t=0.68\,{\rm Myr}$. The three stars marked {\bf A} have already been overrun by the ionisation front, and so their accretion has been terminated. The close binary system marked {\bf B} is an EGG, just emerging from the head of its nascent pillar; its accretion rate is rapidly declining. The star marked {\bf C} is still embedded in the pillar, and growing by accretion; at the end, it is the most massive star.

For higher ionising fluxes, a fragmented bundle of smaller pillars is formed. The individual pillars do not contain sufficient mass to spawn many additional stars, and they are quickly dispersed by the ionisation front.

\subsubsection{Star formation efficiency}\label{SEC:EFF}%

\begin{figure}[h]
\centering
\includegraphics[width=0.8\textwidth]{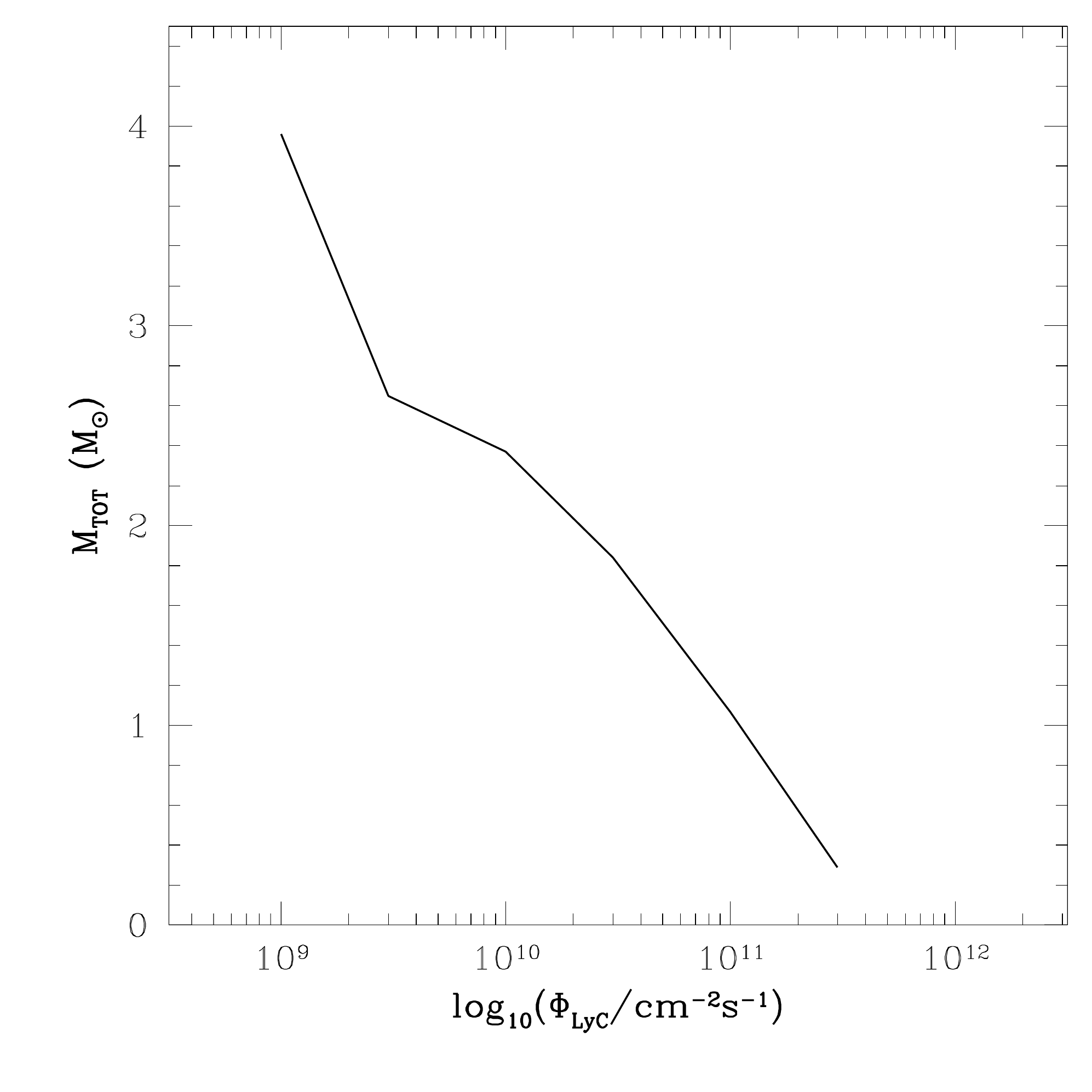}
\caption{Log-linear plot of the ionising flux, $\Phi_{_{\rm LyC}}$ against the total mass of stars formed, $M_{_{\rm TOT}}$, from the fiducial cloud in Simulations 1 to 6.}
\label{FIG:MSTARTOT}
\end{figure}

\begin{figure}[h]
\centering
\includegraphics[width=0.8\textwidth]{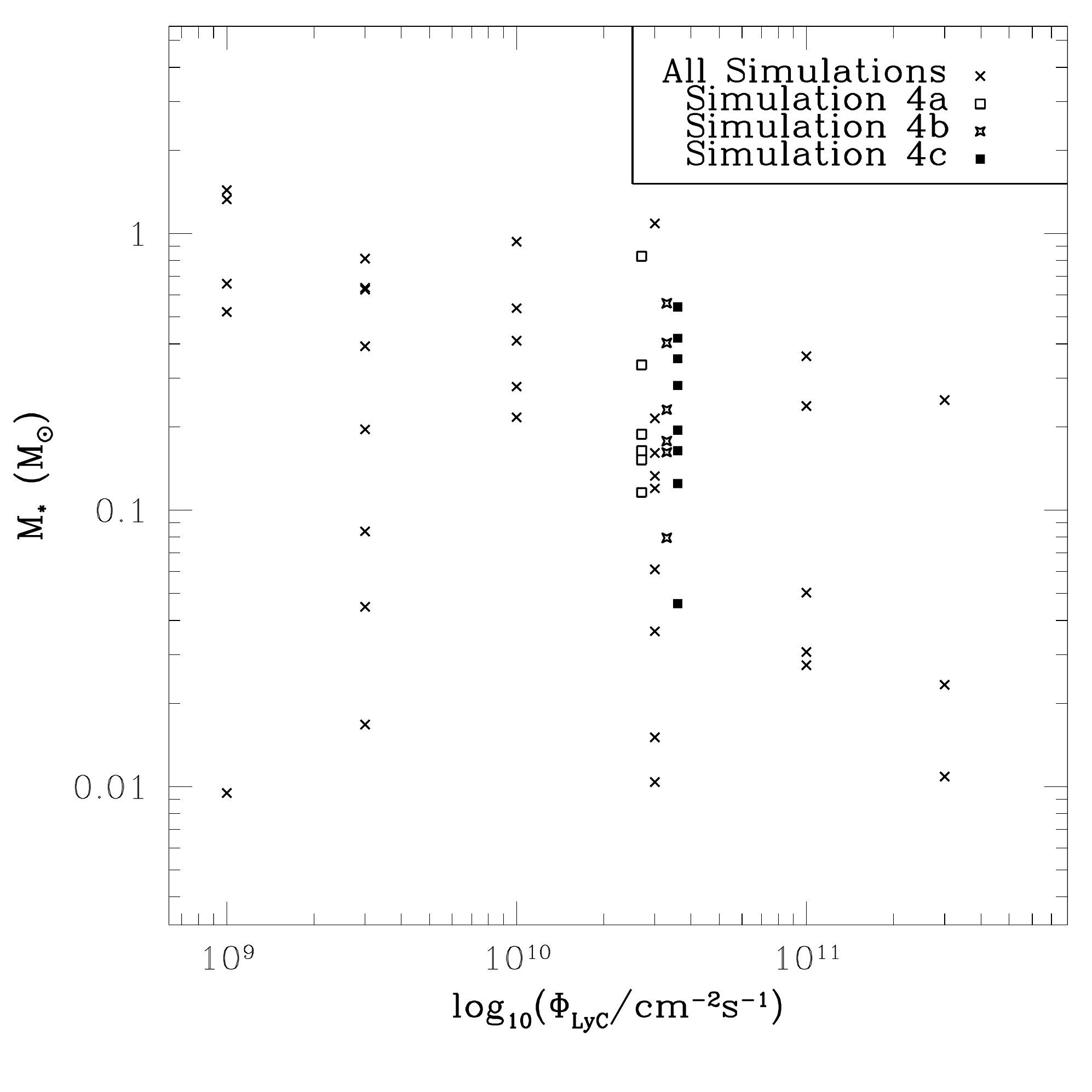}
\caption{Log-log plot of the ionising flux, $\Phi_{_{\rm LyC}}$ against the masses, $M_{_\star}$, of the individual stars formed from the fiducial cloud in Simulations 1 to 6. For Simulation 4 we have also plotted -- at slightly displaced positions -- the masses of the stars formed in the three additional simulations performed to evaluate the influence of numerical noise (Simulations 4a, 4b and 4c; see Section \ref{SEC:NUMNOISE}); because of the intrinsic variance in the outcomes of these different realisations, this produces a small bulge in the upper envelope on the individual masses.}
\label{FIG:MSTARIND}
\end{figure}

Figure~\ref{FIG:MSTARTOT} shows that the total mass of stars, $M_{_{\rm TOT}}$, formed from the fiducial cloud is a monotonically decreasing function of the ionising flux, $\Phi_{_{\rm LyC}}$. Figure~\ref{FIG:MSTARIND} shows that a range of stellar masses is formed, from $0.01\,{\rm M}_{_\odot}$ to $1.44\,{\rm M}_{_\odot}$, with the upper envelope on the individual masses being a decreasing function of $\Phi_{_{\rm LyC}}$.

When the ionising flux is low, the evolution of the cloud is relatively slow. Consequently there is more time for stars to form, and -- once they have formed -- to accrete, because they remain embedded longer. As a result the total mass converted into stars is large ($\sim 0.8M_{_{\rm CLOUD}}$ in Simulation 1), and at the end there are some relatively high-mass stars.

When the ionising flux is high, the evolution of the cloud is relatively rapid, and a large fraction of the mass is eroded by the ionisation front before it can become dense enough to collapse and form stars. Moreover, stars that do form enjoy only a brief embedded phase during which they can grow by accretion. The total mass converted into stars is therefore small ($\sim 0.06M_{_{\rm CLOUD}}$ in Simulation 6), and the individual stellar masses are low.

\begin{figure}[h]
\centering
\includegraphics[width=0.8\textwidth,angle=-90]{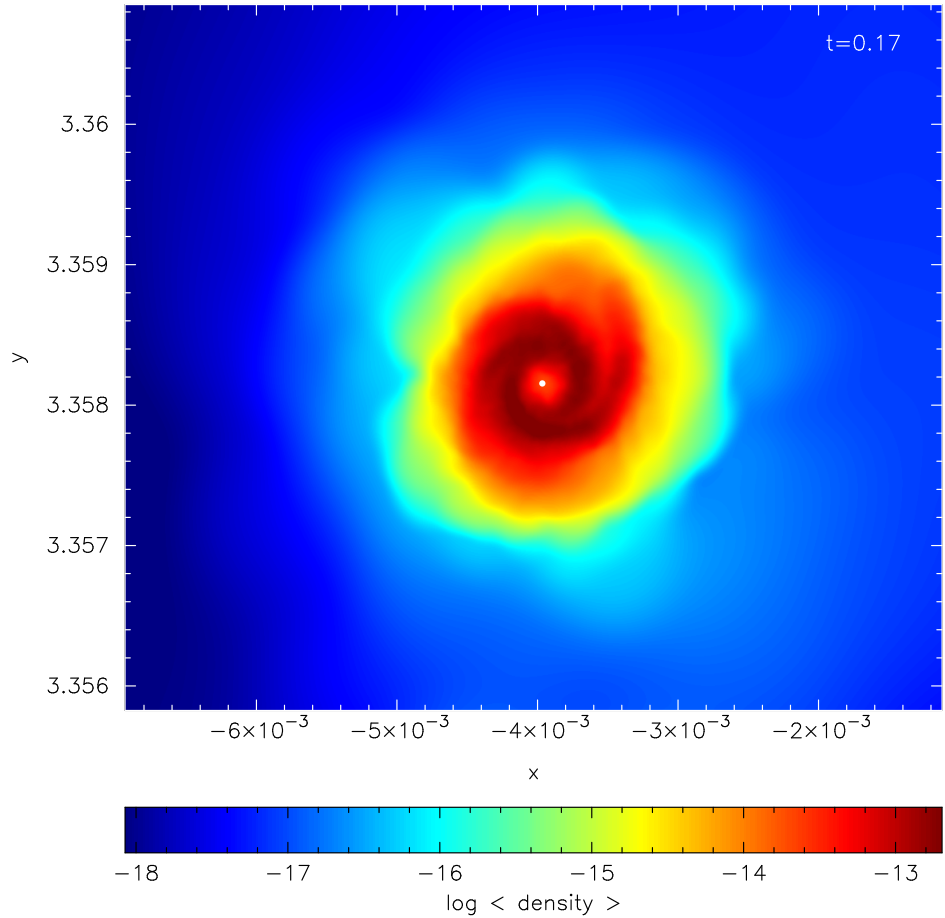}
\caption{Surface-density image of a protostellar disc, taken from Simulation 3 at time $t=0.17\,{\rm Myr}$. See Section \ref{SEC:EFF} for commentary. The colour bar gives $\log_{_{10}}(\Sigma/{\rm g}\,{\rm cm}^{-2})$, where $\Sigma$ is the surface-density, and the axes are in pc.}
\label{FIG:DISC}
\end{figure}

There are too few simulations here to infer more detailed constraints on the statistics of stars formed by radiatively driven implosion, but extended protostellar discs, with radii $\ga 100\,{\rm AU}$, appear to be common, and there are several multiple systems. Figure~\ref{FIG:DISC} shows a face-on projection of a typical protostellar disc, taken from Simulation 3. The star is represented by the white dot with diameter $10\,{\rm AU}$, at the centre of the disc --- although the corresponding sink is actually half this size, with diameter $5\,{\rm AU}$. At this juncture, the star is in a wide binary system with separation $\,\sim\! 1000\,{\rm AU}$. This binary subsequently hardens, due to dissipation between the circumstellar discs attending its individual components, and at the same time it moves towards a third star and forms an hierarchical triple.

\subsection{Non-fiducial clouds exposed to different ionising fluxes}\label{SEC:FLUXMASS}%

\begin{table}
\begin{center}
\caption{The ID, cloud mass, $M_{_{\rm CLOUD}}$, cloud radius, $R_{_{\rm CLOUD}}$, distance from ionising star to cloud, $D_{_{\rm CLOUD}}$, output of ionising photons, $\dot{\cal N}_{_{\rm LyC}}$, and incident ionising flux, $\Phi_{_{\rm LyC}}$, for the non-fiducial clouds simulated in order to generate Figure~\ref{FIG:FLUXMASS}.}
\label{TAB:SUITE2}
\begin{tabular}{ccccccc}
\tableline\tableline
Simulation ID & 9 & 10 & 11 & 12 & 13 & 14 \\
\tableline
$M_{_{\rm CLOUD}}/{\rm M}_{\odot}$ & \multicolumn{2}{c}{$\stackrel{}{2.5}\;\;$} & \multicolumn{2}{c}{10} & \multicolumn{2}{c}{15} \\
$R_{_{\rm CLOUD}}/{\rm pc}$ & \multicolumn{2}{c}{$\stackrel{}{0.15}\;\;$} & \multicolumn{2}{c}{$\stackrel{}{0.60}$} & \multicolumn{2}{c}{$\stackrel{}{0.90}$} \\
$D_{_{\rm CLOUD}}/{\rm pc}$ & \multicolumn{2}{c}{$\stackrel{}{1.5}\;\;$} & \multicolumn{2}{c}{$\stackrel{}{6.0}$} & \multicolumn{2}{c}{$\stackrel{}{9.0}$} \\
$\dot{\cal N}_{_{\rm Lyc}}/{\rm s}^{-1}$ & $\stackrel{}{8.1\times10^{50}}$ & $2.7\times10^{51}$ & $1.3\times 10^{50}$ & $4.3\times 10^{50}$ & $9.7\times10^{49}$ & $2.9\times10^{50}$ \\
$\Phi_{_{\rm LyC}}/{\rm cm}^{-2}{\rm s}^{-1}$ & $3\times10^{12}$ & $10^{13}$ & $3\times 10^{10}$ & $10^{11}$ & $10^{10}$ & $3\times10^{10}$ \\
\tableline
\end{tabular}
\end{center}
\end{table}

In the second suite of simulations, we treat clouds with different masses from the fiducial cloud, namely $M_{_{\rm CLOUD}}=2.5\,{\rm M}_{_\odot},\;\,10\,{\rm M}_{_\odot}\;\,{\rm and}\;\,15\,{\rm M}_{_\odot}$, but the same isothermal sound speed, $a_{_{\rm O}}=0.2\,{\rm km}\,{\rm s}^{-1}$, the same Bonnor-Ebert parameter, $\xi_{_{\rm BE}}=4$, and the same flux divergence $R_{_{\rm CLOUD}}/D_{_{\rm CLOUD}}=0.1$, hence radii $R_{_{\rm CLOUD}}=0.15\,{\rm pc},\;\,0.6\,{\rm pc}\;\,{\rm and}\;\,0.9\,{\rm pc}$ and distances $D_{_{\rm CLOUD}}=1.5\,{\rm pc},\;\,6\,{\rm pc}\;\,{\rm and}\;\,9\,{\rm pc}$, respectively. Here we are only concerned with locating the separatrix between ionising fluxes that do trigger star formation, and those that do not. Therefore, for each cloud mass we only report the results for two ionising fluxes (one either side of the separatrix), and we terminate the simulations either at $t_{_\star}$, when the first star forms, or when only $20\%$ of the cloud mass remains. Table \ref{TAB:SUITE2} gives the ID, $M_{_{\rm CLOUD}}$, $R_{_{\rm CLOUD}}$, $D_{_{\rm CLOUD}}$, $\dot{\cal N}_{_{\rm LyC}}$ and $\Phi_{_{\rm LyC}}$, for the simulations in this suite (Simulations 9 to 14).

\begin{figure}[h]
\centering
\includegraphics[width=0.8\textwidth]{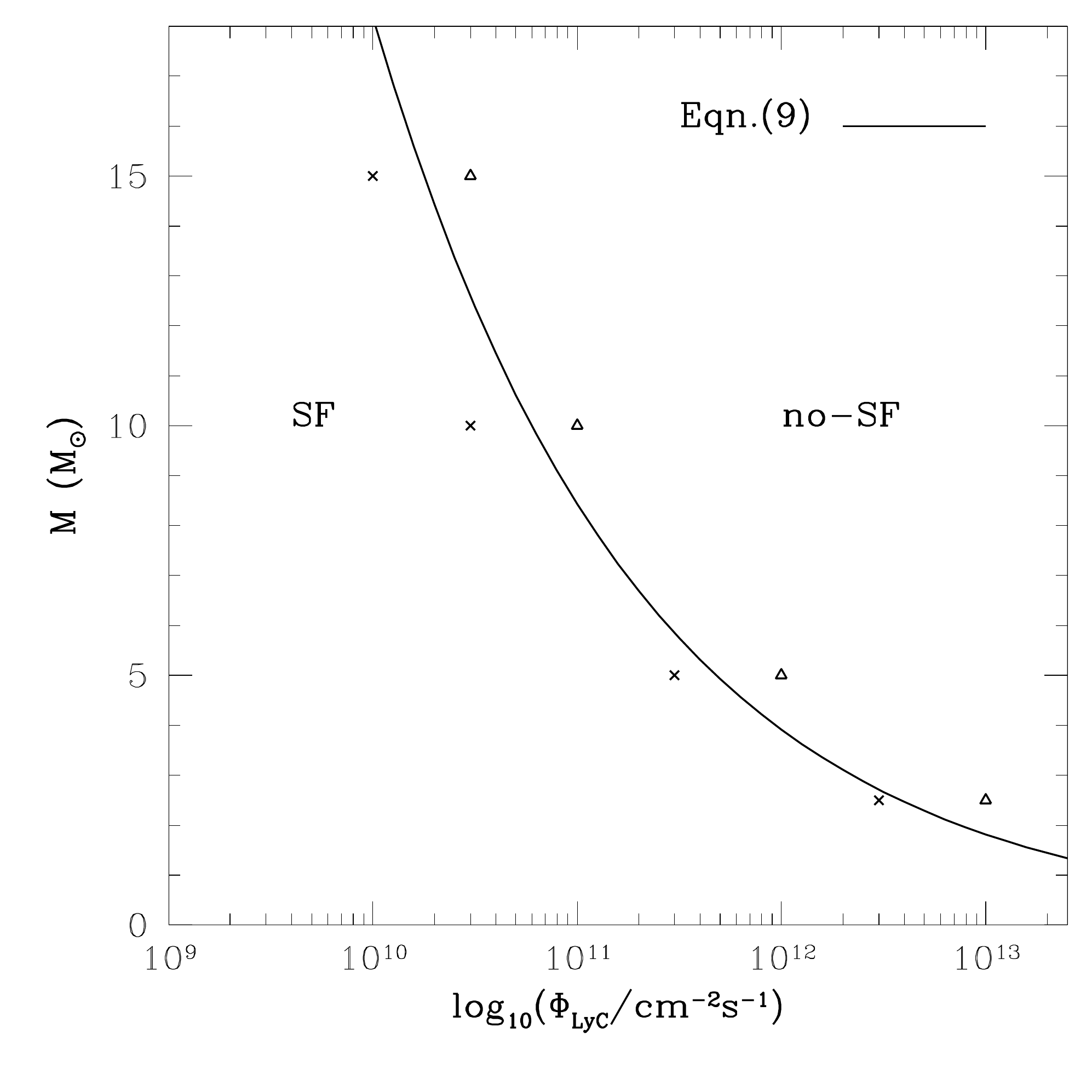}
\caption{Log-linear plot of ionising flux, $\Phi_{_{\rm LyC}}$ against cloud mass, $M_{_{\rm CLOUD}}$ locating the separatrix between configurations that produce stars (SF) and those that do not (no-SF). All other critical parameters are invariant, viz. isothermal sound speed in the cloud, $a_{_{\rm O}}=0.2\,{\rm km}\,{\rm s}^{-1}$, Bonnor-Ebert parameter, $\xi_{_{\rm BE}}=4$, and flux divergence, $R_{_{\rm CLOUD}}/D_{_{\rm CLOUD}}=0.1$. The crosses mark configurations that result in star formation, and the open triangles mark configurations that do not. The solid line shows the locus predicted by Eqn. (\ref{EQN:ANT}).}
\label{FIG:FLUXMASS}
\end{figure}

Figure~\ref{FIG:FLUXMASS} shows, for each of four cloud masses, pairs of ionising fluxes separated by 0.5 dex, for which the lower flux (marked with a cross) triggers star formation and the upper one (marked with an open triangle) does not, but simply disperses the cloud. This plot shows that it requires a higher flux to disperse a less massive cloud without also triggering star formation. This is because the emission measure through a Bonnor-Ebert sphere with fixed isothermal sound speed and fixed Bonnor-Ebert parameter, is proportional to $M_{_{\rm CLOUD}}^{-3}$. Consequently, a less massive cloud is denser, and only a very high ionising flux can penetrate right through it before the ionisation front switches to D-type and starts compressing it.

In support of this contention, we also plot on Figure~\ref{FIG:FLUXMASS} the critical ionising flux for which the ionisation front has just reached the central, approximately uniform-density part of the cloud ($\xi<1$) when it switches to D-type. This is obtained by evaluating the emission-measure integral
\begin{eqnarray}
\zeta(\xi)&=&\int\limits_{\xi'=0}^{\xi'=\xi}\,{\rm e}^{-2\psi(\xi')}\,d\xi'\,,
\end{eqnarray} 
where $\psi(\xi)=\ln(\rho/\rho_{_{\rm CENTRE}})$ is the isothermal function \citep{Cha49}. The critical ionising flux is then
\begin{eqnarray}\label{EQN:ANT}
\Phi_{_{\rm CRIT}}&=&\frac{\alpha_{_{\rm B}}\,a_{_{\rm O}}^{10}\,\mu^3(\xi_{_{\rm BE}})\,\left[\zeta(\xi_{_{\rm BE}})-\zeta(1)\right]}{G^5\,[4\pi m]^2\,M_{_{\rm CLOUD}}^3}\\\label{EQN:NAT}
&\rightarrow&6\times 10^{13}\,{\rm cm}^{-2}\,{\rm s}^{-1}\,\left(\frac{M_{_{\rm CLOUD}}}{{\rm M}_{_\odot}}\right)^{-3},
\end{eqnarray}
where the final expression (Eqn. \ref{EQN:NAT}) has been evaluated for the fiducial parameters $a_{_{\rm O}}=0.2\,{\rm km}\,{\rm s}^{-1}$ and $\xi_{_{\rm BE}}=4$. For ionising fluxes in excess of $\Phi_{_{\rm CRIT}}$, the cloud is simply dispersed, without any star formation being triggered.

\subsection{Influence of numerical noise}\label{SEC:NUMNOISE}%

\begin{table}
\begin{center}
\caption{The ID; time at which star formation starts, $t_{_\star}$; lateral width of the cloud, $w_{_\star}$, when star formation starts; distance behind the ionsation front, $d_{_\star}$, at which star formation first occurs; and total mass of stars formed, $M_{_{\rm TOT}}$, for Simulation 4 and the three additional simulations (4a,4b,4c) performed with different realisations of the fiducial cloud exposed to ionising flux $\Phi_{_{\rm LyC}}=3\times 10^{10}\,{\rm cm}^{-2}\,{\rm s}^{-1}$.}
\label{TAB:SUITE3}      
\centering                          
\begin{tabular}{ccccc}
\tableline\tableline
Simulation ID & 4 & 4a & 4b & 4c \\\tableline
$t_{\star}/{\rm Myr}$ & $\stackrel{}{0.06}$ & 0.055 & 0.06 & 0.06 \\
$w_{\star}/{\rm pc}$ & 0.2 & 0.23 & 0.23 & 0.22 \\
$d_{\star}/{\rm pc}$ & $0.029$ & $0.034$ & $0.024$ & $0.027$ \\
$M_{_{\rm TOT}}/{\rm M}_{\odot}$ & 1.84 & 1.93 & 1.61 & 2.12 \\
\tableline
\end{tabular}
\end{center}
\end{table}

In the third suite of simulations, we treat three additional realisations of the simulation involving the fiducial cloud and an ionising flux $\Phi_{_{\rm LyC}}=3\times 10^{10}\,{\rm cm}^{-2}\,{\rm s}^{-1}$. These simulations are identical to Simulation 4 (presented in Section \ref{SEC:SUITE1}), except that in each case the distribution of SPH particles representing the cloud has been rotated through three random angles. Consequently, the particle noise -- which influences the details of when and where star formation occurs, and therefore what the final stellar masses are -- is different.

\begin{figure}[h]
\centering
\includegraphics[angle=-90,width=0.7\textwidth]{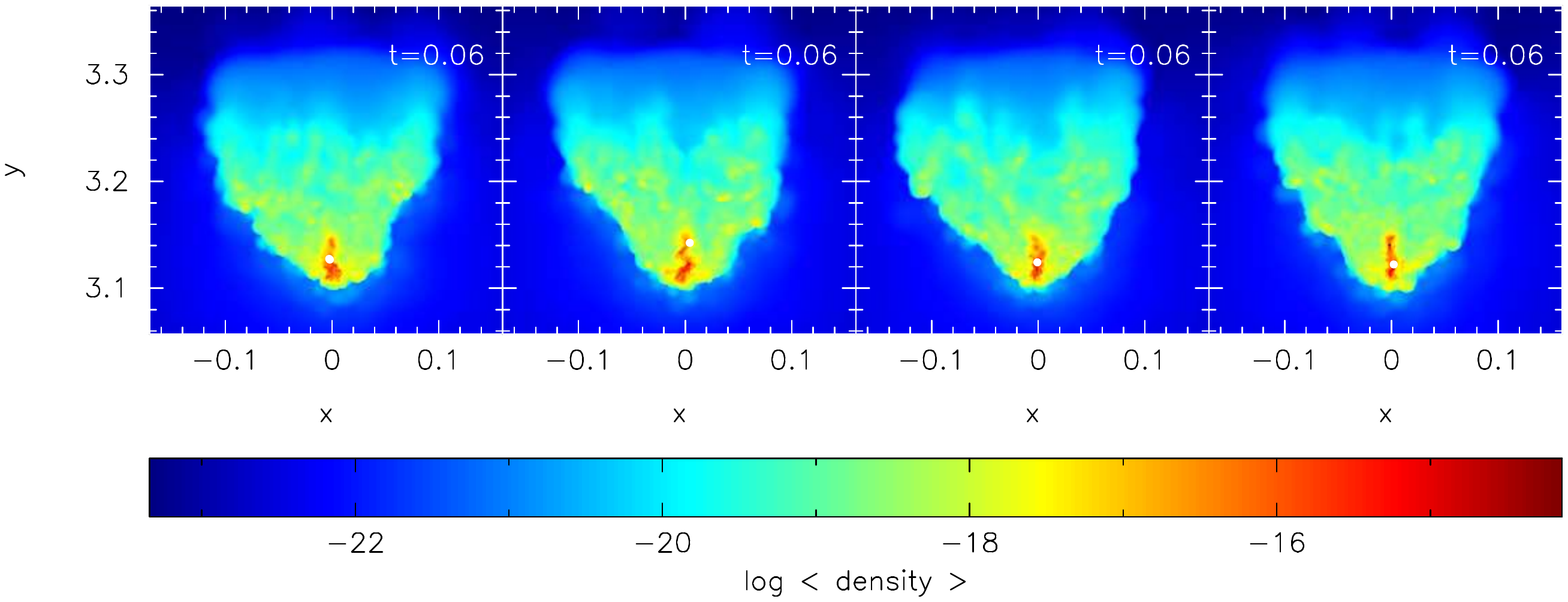}
\includegraphics[angle=-90,width=0.7\textwidth]{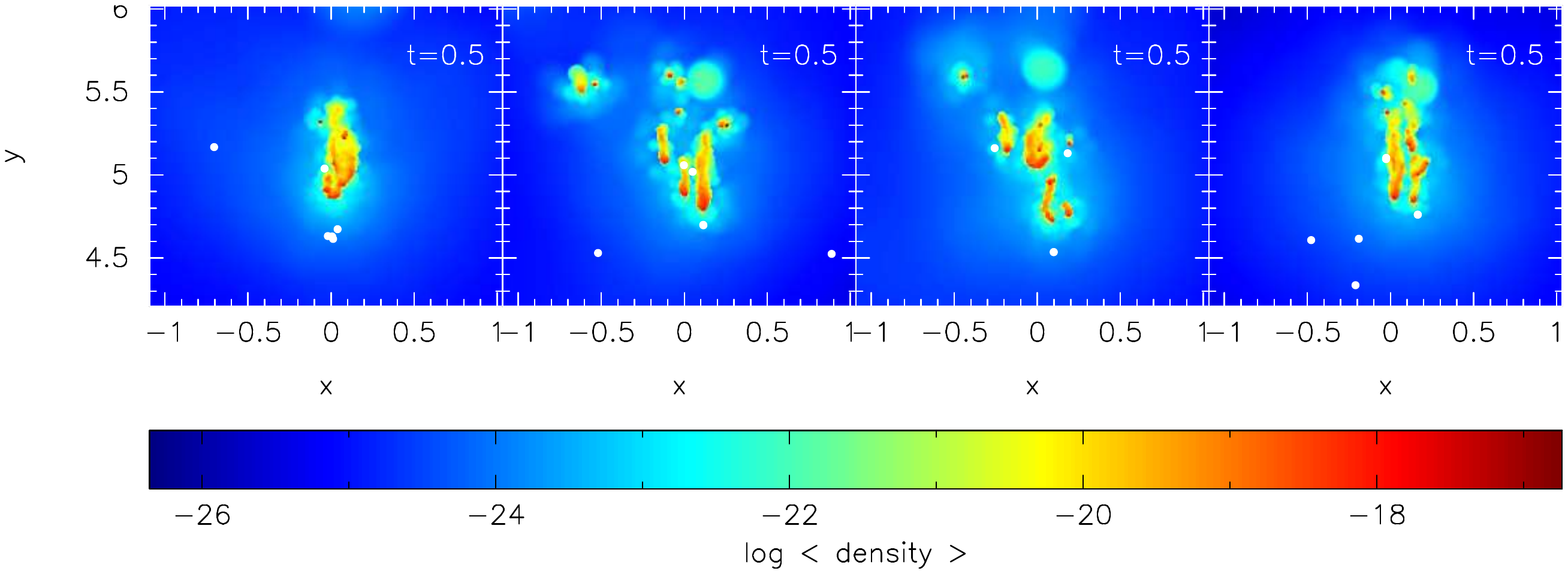}
\caption{Surface-density images of Simulations 4, 4a, 4b and 4c. The first row is at $t_{_\star}=0.06\,{\rm Myr}$ when star formation starts. The cloud morphology is broadly similar in all four simulations, and star formation is located at approximately the same distance $d_{\star}$ behind the ionisation front. The second row is at $t=0.5\,{\rm Myr}$, and the distribution of newly-formed stars and residual neutral gas is more chaotic, but statistically similar. The colour bar gives $\log_{_{10}}(\Sigma/{\rm g}\,{\rm cm}^{-2})$, where $\Sigma$ is the surface-density, and the axes are in pc.}
\label{FIG:SUITE3}
\end{figure}

Table \ref{TAB:SUITE3} summarises the global metrics for Simulations 4, 4a, 4b and 4c; and Figure~\ref{FIG:SUITE3} shows surface-density images for the same simulations at $t_{\star}=0.06\,{\rm Myr}$ (top row) and at $t=0.5\,{\rm Myr}$ (bottom row). All the global metrics ($t_{\star}$, $d_{\star}$, $w_{\star}$ and $M_{_{\rm TOT}}$) are broadly similar, as is the morphology of the cloud at $t_{_\star}$ and the location of star formation. However, as would be expected for a chaotic non-linear process, the details of the final state (masses and locations of individual stars at $t=0.5\,{\rm Myr}$) show a large spread, and this should be interpreted as a measure of the intrinsic variance of the process.

\section{Conclusions}\label{SEC:CONC}%

We present the results of simulations of star formation triggered by radiatively driven implosion, i.e. stable clouds which are driven into collapse and fragmentation by being overrun by an \HIIR. The clouds are all modelled as stable Bonnor Ebert spheres (i.e. equilibrium isothermal spheres) with isothermal sound speed $a_{_{\rm O}}=0.2\,{\rm km}\,{\rm s}^{-1}$ and Bonnor-Ebert parameter $\xi_{_{\rm BE}}=4$. The ionising star is always placed at distance $D_{_{\rm CLOUD}}=10R_{_{\rm CLOUD}}$ from the initial centre of the cloud, but its ionising output, $\dot{\cal N}_{_{\rm LyC}}$, and hence the ionising flux incident on the cloud, $\Phi_{_{\rm LyC}}$, are varied through over three orders of magnitude.

There is a range of ionising fluxes that trigger star formation. In this range, the time at which the first star forms is given approximately by $t_{\star}=0.19\,{\rm Myr}\,\left(\Phi_{_{\rm LyC}}/10^9\,{\rm  cm}^{-2}\,{\rm s}^{-1}\right)^{-1/3}$.

For ionising fluxes at the low end of this range, the evolution of the cloud is quite slow. When star formation starts at $t_{_\star}$, the ionisation front on the exposed side of the cloud has a broad {\bf U}-shape, and star formation is concentrated in a dense filament along the axis of symmetry of the cloud, some distance behind the ionisation front. This finding is in agreement with observations of star formation in bright rimmed clouds, as reported by \citet{Sug99, Sug00}. Although compression of the cloud is relatively weak, star formation is quite efficient (a large fraction of the cloud mass is converted into stars) and the individual stars have relatively high mass, because there is a long period during which star formation can occur; also, once formed, the stars remain embedded in the neutral gas for a long time and can therefore continue to grow by accretion. During the later stages of cloud dispersal, the remnants of the cloud appear as a pillar protruding into the \HIIR; the newly-formed stars emerge from this pillar, and into the \HIIR, as Evaporating Gaseous Globules \citep[EGGs; see][]{Hes96}.

For ionising fluxes at the high end of this range, the evolution of the cloud is more rapid. When star formation starts at $t_{_\star}$, the ionisation front on the exposed side of the cloud has a narrow {\bf V}-shape, and star formation is concentrated at the tip of the {\bf V}, just behind the ionisation front. Although compression of the cloud is relatively strong, star formation is quite inefficient (a small fraction of the cloud mass is converted into stars) and the individual stars have relatively low mass, because there is very limited time available for star formation before the cloud is dispersed by being ionised; also, once formed, the stars do not remain embedded in the neutral gas for long, and so they cannot grow much by accretion. During the later stages of cloud dispersal, the cloud breaks up into a bundle of small pillars which do not have enough mass to spawn further stars. Once the stars emerge into the \HIIR, their growth ceases.

\acknowledgments

The authors would like to thank the referee, Fabrizio Massi, for his useful comments. TGB and RW acknowledge support from the project LC06014-Centre for Theoretical Astrophysics of the Ministry of education, Youth and Sports of the Czech Republic. APW and SW gratefully acknowledge the support of the Marie Curie Research Training Network {\sc CONSTELLATION} (Ref. MRTN-CT-2006-035890). DAH is funded by a Leverhulme Trust Research Project Grand (F/00 118/NJ). The computations in this work were carried out on Merlin Supercomputer of Cardiff University. The data analysis and the column density plots were made using the SPLASH visualization code \citep{Pri07}.

\end{document}